\documentclass[12pt,a4paper]{article}

\usepackage{hyperref}
\hypersetup{
    colorlinks=true,
    linkcolor=blue,
    filecolor=magenta,      
    urlcolor=blue,
    citecolor=blue
}
\usepackage{a4wide}
\usepackage{amsmath}
\usepackage{bm}
\usepackage{amssymb}
\usepackage{graphicx}

\allowdisplaybreaks

\newcommand{\Op}{\mathcal{O}}

\newcommand{\HBS}{\mathbb S}
\newcommand{\BS}{\mathbb S}
\newcommand{\M}{j}

\newcommand{\zt}{\zeta_3}

\newcommand{\zf}{\zeta_5}

\def\z#1{\zeta_#1}
\def\zz#1#2{\zeta_{#1#2}}

\def\hh#1#2{{{{\mathrm h}_{#1#2}}}}

\newcommand{\cN}{{\cal N}}
\newcommand{\sign}{\mathop{\mathrm{sign}}\nolimits}
\newcommand{\Cl}{\mathop{\mathrm{Cl}}\nolimits}
\renewcommand{\thefootnote}{\textit{\alph{footnote}}}

\begin{document}

\thispagestyle{empty}

\vskip-3cm{\baselineskip14pt
\centerline{\rm\normalsize DESY 21-043\hfill ISSN 0418-9833}
\centerline{\rm\normalsize March 2021\hfill}}
\vskip1.5cm
\setcounter{footnote}{0}

\begin{center}
{\Large{\bf
Non-planar universal anomalous dimension\\[3mm] of twist-two operators with general Lorentz spin\\[3mm] at four loops in $\cN=4$ SYM theory
}} \vspace{15mm}

{\sc\bf
B.~A.~Kniehl\footnote{kniehl@desy.de} and V.~N.~Velizhanin\footnote{velizh@thd.pnpi.spb.ru}}\\[8mm]

${}^a${\it Institut f{\"u}r Theoretische Physik,\\
Hamburg Universit{\"a}t,\\
Luruper Chaussee 149,\\
Hamburg, Germany
}\\[5mm]
${}^b${\it Theoretical Physics Division\\
Petersburg Nuclear Physics Institute\\
Orlova Roscha, Gatchina\\
188300 St.~Petersburg, Russia}\\[5mm]

\textbf{Abstract}\\[2mm]
\end{center}

\noindent{
We compute the non-planar contribution to the universal anomalous dimension of twist-two operators in $\cN=4$ supersymmetric Yang-Mills theory at four loops through Lorentz spin eighteen. Exploiting the results of this and our previous calculations along with recent analytic results for the cusp anomalous dimension and some expected analytic properties, we reconstruct a general expression valid for arbitrary Lorentz spin. We study various properties of this general result, such as its large-spin limit, its small-$x$ limit, and others. In particular, we present a prediction for the non-planar contribution to the anomalous dimension of the single-magnon operator in the $\beta$-deformed version of the theory.
}
\newpage

\renewcommand*{\thefootnote}{\fnsymbol{footnote}}
\setcounter{footnote}{1}
\setcounter{page}{1}

\tableofcontents

\section{Introduction}

In the framework of Quantum Chromodynamics (QCD), the anomalous dimensions of composite operators provide us with information on the evolution of the parton distribution functions of the proton through the Dokshitzer-Gribov-Lipatov-Altarelli-Parisi (DGLAP)~\cite{Gribov:1972ri,Gribov:1972rt,Altarelli:1977zs,Dokshitzer:1977sg} equations. The simplest way to obtain these quantities is the direct computation of the renormalization of the composite operators which appear in the operator product expansion of two currents in the framework of electron-proton deep-inelastic scattering. In QCD, this is usually done in this way. At the present time, such calculations are performed analytically through the next-to-next-to-next-to-leading order, i.e., through fourth order in the strong-coupling constant or, equivalently, through four loops in the corresponding diagrammatic technique~\cite{Gross:1973ju,Georgi:1951sr,Ahmed:1976ee,Floratos:1977au,GonzalezArroyo:1979df,Floratos:1978ny,GonzalezArroyo:1979he,Larin:1991fx,Larin:1993vu,Mertig:1995ny,Retey:2000nq,Moch:2004pa,Vogt:2004mw,Baikov:2006ai,Velizhanin:2011es,Baikov:2015tea,Moch:2017uml,Moch:2018wjh,Velizhanin:2014fua}.

On the other hand, by definition, the anomalous dimension is the quantum correction to the canonical dimension of the product of two (or more) elementary fields or composite operators. This approach is usually adopted in conformal field theories. Both approaches have actively been used after the discovery of the AdS/CFT correspondence~\cite{Maldacena:1997re,Gubser:1998bc,Witten:1998qj} between supergravity in anti-de~Sitter space and four-dimensional conformal field theory, $\mathcal{N}=4$ supersymmetric Yang-Mills (SYM) theory. 
The great interest in calculations of anomalous dimensions of composite operators comes from investigations of integrability in the framework of the AdS/CFT correspondence. In the planar limit, both direct computations~\cite{Lipatov:1997vu,Anselmi:1998ms,Bianchi:2000hn,Lipatov:2000,Arutyunov:2001mh,Dolan:2001tt,Kotikov:2003fb,Kotikov:2004er,Eden:2004ua,Bern:2006ew,Kotikov:2007cy,Fiamberti:2007rj,Fiamberti:2008sh,Velizhanin:2008jd,Velizhanin:2008pc} and computations via the generalized L\"uscher corrections~\cite{Luscher:1985dn,Luscher:1986pf} were performed~\cite{Bajnok:2008bm,Bajnok:2008qj,Beccaria:2009eq,Bajnok:2009vm,Lukowski:2009ce,Velizhanin:2010cm,Bajnok:2012bz} at higher orders of perturbation theory to test the Asymptotic Bethe Ansatz~\cite{Minahan:2002ve,Beisert:2003tq,Beisert:2003yb,Dolan:2003uh,Bena:2003wd,Kazakov:2004qf,Beisert:2004hm,Arutyunov:2004vx,Staudacher:2004tk,Beisert:2005di,Beisert:2005bm,Beisert:2005fw,Beisert:2005cw,Janik:2006dc,Hernandez:2006tk,Arutyunov:2006iu,Beisert:2006ib,Beisert:2006ez,Beisert:2007hz,Beisert:2010jr} as well as 
the Quantum Spectral Curve relations~\cite{Arutyunov:2009zu,Gromov:2009tv,Arutyunov:2009ur,Bombardelli:2009ns,Gromov:2009bc,Arutyunov:2009ax,Gromov:2013pga,Gromov:2014caa}. 
The latter allowed one to compute the general result for the anomalous dimension of the twist-two operators through seven loops~\cite{Marboe:2014sya,Marboe:2016igj} and, for special values of the Lorentz spin $j$, even through eleven loops~\cite{Leurent:2012ab,Leurent:2013mr,Marboe:2014gma,Marboe:2018ugv}.

The computation of the non-planar contribution to the universal anomalous dimension of the twist-two operators in $\mathcal{N}=4$ SYM theory, which is suppressed by the inverse color factor $N_c$ of the gauge group, is of great interest 
in the context of gauge/string duality, as it is related to the loop corrections of string amplitudes. 
Information on non-planar corrections can be obtained by direct computation of the appropriate non-planar Feynman diagrams, via advanced computerized methods of four-loop calculation. Such computations were performed in Refs.~\cite{Velizhanin:2009gv,Velizhanin:2010ey,Velizhanin:2014zla}. Alternatively, they can be calculated by applying the method of asymptotic expansion to the four-point functions of length-two half-Bogomol'nyi-Prasad-Sommerfield operators~\cite{Fleury:2019ydf}.
These results allows us deepen our understanding of the AdS/CFT correspondence in regions already accessed and to explore new ones.
Having the general result for the non-planar part of the general anomalous dimension of the twist-two operators, for general value of their Lorentz spin $j$, allows us to study its particular limits. The most interesting one, of large Lorentz spin $j$, yields the cusp anomalous dimension, which can also be computed by other methods and which, in $\mathcal{N}=4$ SYM theory, was obtained to all orders in the planar limit~\cite{Beisert:2006ez}.
Recently, its non-planar part has been established through four loops, at
$\mathcal{O}(g^8)$, via the Sudakov form factor, numerically in
Refs.~\cite{Boels:2015yna,Boels:2017skl} and analytically in
Ref.~\cite{Huber:2019fxe}, and via light-like polygonal Wilson loops, analytically in Ref.~\cite{Henn:2019swt}.
At four loops in QCD, at $\mathcal{O}(\alpha_s^4)$ in the strong-coupling
constant $\alpha_s$, the quark cusp anomalous dimension in the planar limit was
found via the quark form factor in Ref.~\cite{Lee:2016ixa}, its
contribution with quartic fundamental color factor was obtained, again via
the quark form factor, in Ref.~\cite{Lee:2019zop}, and the complete quark and
gluon cusp anomalous dimensions were established via their counterpart in
${\mathcal N}=4$ SYM theory in Ref.~\cite{Henn:2019swt} and via the massless
quark and gluon form factors in Ref.~\cite{vonManteuffel:2020vjv}.
Moreover, it is of interest to study modifications of the Balitsky--Fadin--Kuraev--Lipatov (BFKL)~\cite{Lipatov:1976zz,Kuraev:1977fs,Balitsky:1978ic} and the double-logarithmic~\cite{Kirschner:1982qf,Kirschner:1983di} equations due to non-planar contributions. The latter can easily be obtained from the general result by means of analytic continuation.

As for the universal anomalous dimension of the twist-two operators in $\mathcal{N}=4$ SYM theory, the results previously obtained~\cite{Velizhanin:2009gv,Velizhanin:2010ey,Velizhanin:2014zla} for the first three nontrivial even values of $j$, for $j=4,6,8$, gave hope for the feasibility to reconstruct the general result, for generic value of $j$, by means of special methods based on number theory. In fact, this provides a strong motivation for us to proceed to higher values of $j$. In a recent letter~\cite{Kniehl:2020rip}, we summarized our new results for the next five nontrivial $j$ values, $j=10,\ldots,18$, and performed a completely independent numerical computation of the cusp anomalous dimensions.
In the following, we provide full details of the calculation and, on top of that, conjecture an analytic expression for arbitrary value of $j$.

\boldmath
\section{Computations for fixed values of $j$}
\unboldmath

We consider the following SU(4)-singlet, twist-two operators within $\mathcal{N}=4$ SYM theory:
\begin{eqnarray}
\mathcal{O}_{\mu _{1},...,\mu _{\M}}^{\lambda } &=&\hat{S}
\bar{\lambda}_{i}^{a}\gamma _{\mu _{1}}
    {\mathcal D}_{\mu _{2}}\cdots{\mathcal D}_{\mu _{\M}}\lambda ^{a\;i}\,, \label{qqs}
    \\
\mathcal{O}_{\mu _{1},...,\mu _{\M}}^{g} &=&\hat{S} G_{\rho \mu_{1}}^{a}{\mathcal
D}_{\mu _{2}} {\mathcal D}_{\mu _{3}}\cdots{\mathcal D}_{\mu _{\M-1}}G_{\phantom{a,\rho}\mu_{\M}}^{a,\rho}\,,
\label{ggs}\\
\mathcal{O}_{\mu _{1},...,\mu _{\M}}^{\phi } &=&\hat{S}
\bar{\phi}_{r}^{a}{\mathcal D}_{\mu _{1}} {\mathcal D}_{\mu _{2}}\cdots{\mathcal
D}_{\mu _{\M}}\phi _{r}^{a}\,,\label{phphs}
\end{eqnarray}
where ${\mathcal D}_{\mu_i}$ denotes the covariant derivative, the spinors $\lambda_{i}$ and the field-strength tensor $G_{\rho \mu }$ describe gauginos and gauge fields, respectively, and $\phi_{r}$ are the complex scalar fields appearing in ${\mathcal N}=4$ SYM theory. The indices $i=1,2,3,4$ and $r=1,2,3$ refer to the SU(4) and SO(6)${}\simeq{}$SU(4) groups of inner symmetry, respectively. The symbol $\hat{S}$ implies a symmetrization of each tensor in the Lorentz indices $\mu_{1},...,\mu_{\M}$ and a subtraction of its traces. The operator in Eq.~\eqref{qqs} is familiar from the same type of computations in QCD.
The operators in Eqs.~\eqref{qqs}--\eqref{phphs} form the multiplicatively renormalized operators.
Their anomalous dimensions are expressed through the so-called universal anomalous dimension,
\begin{equation}
  \gamma_{\mathrm {uni}}(\M)\ =\ \sum_{n=1}^{\infty}\gamma_{\mathrm {uni}}^{(n-1)}(\M)\,g^{2n}\,,
  \label{eq:gamuni}
\end{equation}
where $g^2=\lambda/(16\pi^2)$ with $\lambda=g^2_{\mathrm {YM}}N_c$ being the 't~Hooft coupling constant, up to integer argument shifts \cite{Kotikov:2002ab}.
In particular, the universal anomalous dimension $\gamma_{\mathrm{uni}}(j)$ in Eq.~\eqref{eq:gamuni} is related to the anomalous dimension of the frequently studied twist-two operator 
\begin{equation}
\mathcal{O}^M_{\mathcal{Z}}=\mathrm {Tr} \mathcal{Z}{\mathcal D}_{\mu _{1}} {\mathcal D}_{\mu _{2}}\cdots{\mathcal
D}_{\mu _{M}}\mathcal{Z}\,,
\end{equation}
where $\mathcal{Z}$ is one of the scalar fields $\phi_r$, which belong to the $\mathfrak{sl}(2)$ sub-sector of $\mathcal{N}=4$ SYM theory, just by shifting the argument, as
\begin{equation}
\gamma_{\mathrm{uni}}(j)=\gamma_{\mathcal{O}^M_{\mathcal{Z}}}(M+2)\,.
\end{equation}

Non-planar contributions arise from the Feynman diagrams which contain the following combinations of color structures:
\begin{equation}
d^{abcd} =\frac{1} {6}\left[
\mathrm{Tr}\left(f^{paq}f^{qbr}f^{rcs}f^{sdp}\right)
 + \mathrm{five}\   bcd\ \mathrm{permutations}\right]\,.
\end{equation}
It hence follows that
\begin{equation}
d_{44}=d^{abcd}d_{abcd}=\frac{1}{24}N_c^2(N_c^2+36)=
{N_c^4}\left(\frac{1}{24}+\frac{3}{2}\frac{1}{N_c^2}\right)
=
\frac{{N_c^4}}{32}\left(\frac{4}{3}+\frac{48}{N_c^2}\right)\,.
\end{equation}
From Eq.~\eqref{eq:gamuni} we thus glean that the non-planar (np) four-loop ($n=4$) contributions of Lorentz spin $j$ to the universal anomalous dimension,
$\gamma_{\mathrm{uni,np}}^{(3)}(j)$, are proportional to $g^8/N_c^2$.
For the reader's convenience, we recall here the previously computed results
for $j=4,6,8$ \cite{Velizhanin:2009gv,Velizhanin:2010ey,Velizhanin:2014zla}: 
\begin{eqnarray}
\gamma_{\mathrm {uni,np}}^{(3)}(4)&=&
- 360\zf\, \frac{48}{N_c^2}
\,,
\label{guniM4}\\
\gamma_{\mathrm {uni,np}}^{(3)}(6)&= &
\frac{25}{9}\left(21 + 70 \zt - 250 \zf \right)  \frac{48}{N_c^2}
\,,
\label{guniM6}\\
\gamma_{\mathrm {uni,np}}^{(3)}(8)& =&
\frac{49}{600} \left(1357 + 4340 \zt - 11760 \zf\right)  \frac{48}{N_c^2}\,,
\label{guniM8}
\end{eqnarray}
where $\zeta_n=\zeta(n)$ is Riemann's zeta function.
We observe that there are rather simple common factors on the right-hand sides of Eqs.~\eqref{guniM4}--\eqref{guniM8}, which we pulled out.
In fact, the prefactors in Eqs.~\eqref{guniM4}--\eqref{guniM8} resemble the
harmonic sums $\sum_{i=1}^{j-2}\frac{1}{i}$ for $j=4,6,8$, with values $3/2$,
$25/12$, $49/20$, and harmonic sums are also expected to appear as building
blocks of $\gamma_{\mathrm {uni,np}}^{(3)}(j)$, as explained below.
If this kind of factorization were preserved for the higher values of $j$, this would considerably simplify the procedure for finding the general form of the universal anomalous dimension from the results for fixed values of $j$. 

We now proceed to the case of $j=10$, where we encounter two difficulties in the application of the method established in Refs.~\cite{Velizhanin:2009gv,Velizhanin:2010ey,Velizhanin:2014zla}. 
First, it is necessary to extend the database of all Feynman integrals which can be contained in the computed Feynman diagrams. Such a database is formed by reducing the considered Feynman integrals to master integrals. For this purpose, we use our MATHEMATICA realization of the Laporta algorithm~\cite{Laporta:2001dd}. 
Second, a new vertex, with two fermion and four gluon lines, arises from the operator in Eq.~\eqref{qqs} starting at $j=5$. It did not appear in our previous calculations~\cite{Velizhanin:2009gv,Velizhanin:2010ey,Velizhanin:2014zla}, as we exploited properties of the mixing matrix to simplify our computations. This simplification can be easily explained by means of well-known results at the leading order. The matrix of anomalous dimensions for the operators in Eqs.~\eqref{qqs}--\eqref{phphs}, sandwiched between definite states (fermions, gauge field, or scalars), has the following form:
\begin{eqnarray}
\gamma^{(0)}_{{gg}}&=&-4S_1(j)+\frac{4}{j-1}-\frac{4}{j}+\frac{4}{j+1}-\frac{4}{j+2}\,,\qquad
\gamma^{(0)}_{{\lambda g}} \, =\, \frac{8}{j}-\frac{16}{j+1}+\frac{16}{j+2}\,,
\nonumber \\
\gamma^{(0)}_{{\phi g}}&=&\frac{12}{j+1}-\frac{12}{j+2}\,,\qquad
\gamma^{(0)}_{{g\lambda}} \, =\, \frac{4}{j-1}-\frac{4}{j}+\frac{2}{j+1}\,,\qquad
\gamma^{(0)}_{{\lambda\phi}} \, =\, \frac{8}{j}\,,\qquad
\gamma^{(0)}_{{\phi\lambda}} \, =\, \frac{6}{j+1}\,,
\nonumber \\
\gamma^{(0)}_{{\lambda\lambda}} &=&-4S_1(j)+\frac{8}{j}-\frac{8}{j+1}\,,\qquad
\gamma^{(0)}_{{g\phi}} \, =\, \frac{4}{j-1}-\frac{4}{j}\,,\qquad
\gamma^{(0)}_{{\phi \phi}} \, =\, -4 S_1(j)\,.
\label{LOAD}
\end{eqnarray}
The matrix which diagonalizes the matrix of anomalous dimensions can be used for the construction of the following multiplicatively renormalizable operators:
\begin{eqnarray}
\Op^{T_j}_{\mu_1,\ldots,\mu_j} & = & \Op^g_{\mu_1,\ldots,\mu_j} +
\Op^\lambda_{\mu_1,\ldots,\mu_j} + \Op^\phi_{\mu_1,\ldots,\mu_j}\,,
\label{mrop1j}\\
\Op^{\Sigma_j}_{\mu_1,\ldots,\mu_j} & = & - 2(j-1)\Op^g_{\mu_1,\ldots,\mu_j} +
\Op^\lambda_{\mu_1,\ldots,\mu_j} + \frac{2}{3}(j+1)\Op^\phi_{\mu_1,\ldots,\mu_j}\,,
\label{mrop2j}\\
\Op^{\Xi_j}_{\mu_1,\ldots,\mu_j} & = & -\frac{j-1}{j+2}\Op^g_{\mu_1,\ldots,\mu_j}
+ \Op^\lambda_{\mu_1,\ldots,\mu_j} -
\frac{j+1}{j}\Op^\phi_{\mu_1,\ldots,\mu_j}\,,
\label{mrop3j}
\end{eqnarray}
whose anomalous dimensions are $\gamma_{\mathrm {uni}}^{(0)}(j)$, $\gamma_{\mathrm {uni}}^{(0)}(j+2)$, and $\gamma_{\mathrm {uni}}^{(0)}(j+4)$, respectively, where
\begin{equation}
  \gamma_{\mathrm {uni}}^{(0)}(j)=-4S_1(j)\,,
  \label{gammauni0}
\end{equation}
with $S_1(j)=\sum_{i=1}^{j}\frac{1}{i}$ being the simplest harmonic sum.

Sandwiching the operators in Eqs.~\eqref{mrop1j}--\eqref{mrop3j} between different states, we obtain the following set of relations:
\begin{eqnarray}
\gamma_{gg}+      \gamma_{\lambda g}+     \gamma_{\phi g}     &=& \gamma_{\mathrm {uni}}^{(0)}(j)\,,\label{Op1g}\\
\gamma_{g\lambda}+\gamma_{\lambda\lambda}+\gamma_{\phi\lambda} &=& \gamma_{\mathrm {uni}}^{(0)}(j)\,,\label{Op1l}\\
\gamma_{g\phi}+   \gamma_{\lambda\phi}+   \gamma_{\phi\phi}   &=& \gamma_{\mathrm {uni}}^{(0)}(j),\label{Op1s}\\
                 \gamma_{gg}   -\frac{1}{2(j-1)}\gamma_{\lambda g}     +\frac{1}{3}\,\frac{j+1}{j-1}\gamma_{\phi g} &=& \gamma_{\mathrm {uni}}^{(0)}(j+2)\,,\label{Op2g}\\
-2(j-1)\gamma_{g\lambda}+                 \gamma_{\lambda\lambda}+\frac{2}{3}(j+1)          \gamma_{\phi\lambda} &=& \gamma_{\mathrm {uni}}^{(0)}(j+2)\,,\label{Op2l}\\
-3\frac{j-1}{j+1}\gamma_{g\phi}+\frac{3}{2(j+1)}\gamma_{\lambda\phi}   +                          \gamma_{\phi\phi} &=& \gamma_{\mathrm {uni}}^{(0)}(j+2)\,,\label{Op2s}\\
\gamma_{gg}      -\frac{j+2}{j-1}\gamma_{\lambda g}+\frac{j+1}{j}\,\frac{j+2}{j-1}\gamma_{\phi g} &=& \gamma_{\mathrm {uni}}^{(0)}(j+4)\,,\label{Op3g}\\
-\frac{j-1}{j+2}\gamma_{g\lambda}+          \gamma_{\lambda\lambda}-\frac{j+1}{j}               \gamma_{\phi\lambda} &=& \gamma_{\mathrm {uni}}^{(0)}(j+4)\,,\label{Op3l}\\
\frac{j-1}{j+2}\,\frac{j}{j+1}\gamma_{g\phi}-\frac{j}{j+1}\gamma_{\lambda\phi}+                            \gamma_{\phi\phi} &=& \gamma_{\mathrm {uni}}^{(0)}(j+4)\,,\label{Op3s}
\end{eqnarray}
where, on the left-hand sides, diagonal elements are normalized to unity and the arguments of the matrix elements are equal to $j$. 
Thus, if we calculate the anomalous dimension for the operators in Eq.~\eqref{qqs}--\eqref{phphs} at some fixed values of $j$, we obtain the result for the universal anomalous dimension not only for $j$, but also for $(j+2)$ and $(j+4)$. In order to obtain $\gamma_{\mathrm {uni}}(10)$, it is sufficient to compute $\gamma_{\lambda\lambda}(6)$. In fact, the operator in Eq.~\eqref{qqs} for $j=6$ does contain the vertex operator mentioned above. If we proceed to four loops, we are able to compute the contribution to the universal anomalous dimension similarly as in the leading order. The non-planar contribution 
appears for the first time at this order because there are no additional contributions from the renormalization. 
We already exploited this property in our previous four-loop calculations.

Having circumvented both difficulties, we obtain the following result for the anomalous dimension of the operator in Eq.~\eqref{qqs} for $j=6$:
\begin{equation}
\hat\gamma_{\lambda\lambda,\mathrm{np}}^{(3)}(6) =
\left(
\frac{5148948727}{47628000}
-\frac{943}{135}\,{\mathsf{S}}_2
+\frac{2767689751}{8164800}\,\zt
-\frac{2071042}{2205}\,\zf
\right)
\frac{48}{N_c^2}\,,\quad
\label{gllM4op}
\end{equation}
where ${\mathsf{S}}_2=\frac{4}{9\sqrt{3}}\Cl_2(\pi/3)$ \cite{Czakon:2004bu}, with $\Cl_2$ being Clausen's function.
We have yet to add the contribution from the counterterm diagrams due to the non-gauge-invariant massive operator 
$m^2 {\mathcal A}^{\mu_1}{\mathcal A}^{\mu_2}{\mathcal A}^{\mu_3}{\mathcal A}^{\mu_4}$ discussed in Ref.~\cite{Velizhanin:2014zla},
\begin{equation}
\gamma_{\lambda\lambda,\mathrm{ren},\mathrm{np}}^{(3)}(6) =
\left(
-\frac{19099}{38880}
+\frac{943}{135}\,{\mathsf{S}}_2
+\frac{146951}{46656}\,\zt
\right)
 \frac{48}{N_c^2}\,.
\label{gllM4r}
\end{equation}
Combining Eqs.~\eqref{gllM4op} and \eqref{gllM4r}, we obtain
\begin{equation}
\gamma_{\lambda\lambda,\mathrm{np}}^{(3)}(6)=
\left(
\frac{47458819}{441000}
+\frac{1077703}{3150}\,\zt
-\frac{2071042}{2205}\,\zf
\right)\,
  \frac{48}{N_c^2}\,,
\label{gllM4}
\end{equation}
where ${\mathsf{S}}_2$ has canceled out as expected.

Inserting Eq.~\eqref{gllM4} in Eqs.~\eqref{Op1l}, \eqref{Op2l}, and \eqref{Op3l}, we obtain the following expression for the non-planar contribution to the tenth moment of the universal anomalous dimension of the twist-two operators:\footnote{This result was obtained in 2018, but not published.}
\begin{equation}
\gamma_{\mathrm {uni,np}}^{(3)}(10) =
\left(\frac{220854227}{1411200}+\frac{27357}{56}\,\zt-\frac{579121}{490}\,\zf\right)\frac{48}{N_c^2}
\,.\label{guniM10}
\end{equation}
Unfortunately, it is not possible to pull out a common multiplier from this expression, as was possible for Eqs.~\eqref{guniM4}---\eqref{guniM8}. So, our expectation regarding factorization is not confirmed. We note that the results in Eqs.~\eqref{guniM4}--\eqref{guniM8}, and \eqref{guniM10} have recently been obtained using an alternative method in Ref.~\cite{Fleury:2019ydf}, by performing the asymptotic expansion of the four-point functions for length-two half-Bogomol'nyi-Prasad-Sommerfield operators.

For the calculation of the higher moments, we employ the \texttt{FORM}~\cite{Vermaseren:2000nd} package \texttt{FORCER} \cite{Ruijl:2017cxj}, which has recently been developed for computations of this type and was also used to compute anomalous dimensions of twist-two operators in QCD~\cite{Moch:2017uml,Moch:2018wjh}. By means of this tool, we compute the matrix element $\gamma_{\phi\lambda}$, i.e.\ the anomalous dimension of the scalar operator in Eq.~\eqref{phphs} sandwiched between fermionic states, appearing in Eqs.~\eqref{Op1l}, \eqref{Op2l}, and \eqref{Op3l} and, in a similar way as described after Eqs.~\eqref{Op1g}--\eqref{Op3s}, obtain the next four even moments,
\allowdisplaybreaks[1]
\begin{eqnarray}
\gamma_{\mathrm {uni,np}}^{(3)}(12)& =&
\bigg(\frac{28337309747461}{144027072000}
+\frac{345385183}{571536}\,\zt
-\frac{54479161}{39690}\,\zf\bigg)\frac{48}{N_c^2}
\,,\label{guniM12}\\
\gamma_{\mathrm {uni,np}}^{(3)}(14)& =&
\bigg(\frac{9657407179406311}{41493513600000}
+\frac{158654990663}{224532000}\,\zt
-\frac{7399612441}{4802490}\,\zf\bigg)\frac{48}{N_c^2}
\,,
\label{guniM14}\\
\gamma_{\mathrm {uni,np}}^{(3)}(16)& =&
\bigg(\frac{74429504651244877}{280496151936000}
+\frac{205108095887}{256864608}\,\zt
-\frac{1372958223289}{811620810}\,\zf\bigg)\frac{48}{N_c^2}\,,\qquad
\label{guniM16}\\
\gamma_{\mathrm {uni,np}}^{(3)}(18)& =&
\bigg(\frac{8122582838282649980649377}{27516111512617728000000}
+\frac{72169501556777041}{81811377648000}\,\zt\nonumber\\
&&{}-\frac{5936819760481}{3246483240}\,\zf\bigg)\frac{48}{N_c^2}\,.
\label{guniM18}
\end{eqnarray}
Moreover, we reproduce our previous results in Eqs.~\eqref{guniM4}--\eqref{guniM8} and \eqref{guniM10}.

\section{Reconstruction}\label{Reconstruction}

Equipped with Eq.~\eqref{guniM10}--\eqref{guniM18}, we now attempt to reconstruct the general form of the non-planar contribution to the four-loop universal anomalous dimension, i.e.\ to determine the $j$ dependence of the coefficients of $\zeta_5$ and
$\zeta_3$ and the rational reminder in the ansatz 
\begin{equation}\label{HSZ5ResALL}
\gamma_{{\mathrm{uni}},{\mathrm{np}}}^{(3)}(\M)=\left[\gamma_{{\mathrm{uni}},{\mathrm{np}},\zf}^{(3)}(\M)\,\zf+\gamma_{{\mathrm{uni}},{\mathrm{np}},\zt}^{(3)}(\M)\,\zt
+\gamma_{{\mathrm{uni}},{\mathrm{np}},{\mathrm{rational}}}^{(3)}(\M)\right]\frac{48}{N_c^2}\,.
\end{equation}
For this purpose, we use our method for the reconstruction of general results from results for fixed values of $j$ using number theory, which was proposed in Ref.~\cite{Velizhanin:2010cm} and successfully applied to the reconstruction of anomalous dimensions in Refs.~\cite{Velizhanin:2012nm,Velizhanin:2013vla,Marboe:2014sya,Marboe:2016igj}. This method is based on two observations.

First, we assume that we know all the basis functions which the answer contains. For anomalous dimensions in $\mathcal{N}=4$ SYM theory, these are the generalized harmonic sums. They are defined as~\cite{Vermaseren:1998uu,Blumlein:1998if}
\begin{eqnarray} \label{vhs1}
S_a (M)&=&\sum^{M}_{j=1} \frac{[\sign(a)]^{j}}{j^{\vert a\vert}}\,,
\\
S_{a_1,\ldots,a_n}(M)&=&\sum^{M}_{j=1} \frac{[\sign(a_1)]^{j}}{j^{\vert a_1\vert}}
\,S_{a_2,\ldots,a_n}(j)\, ,\label{vhs}
\end{eqnarray}
where the indices $a_1,\ldots,a_n$ may be positive or negative, except for the value $-1$.
The weight or \textit{transcendentality} $\ell$ of each sum $S_{a_1,\ldots,a_n}(M)$ is defined as the sum of the absolute values of its indices,
\begin{equation}
\ell=\vert a_1 \vert +\cdots+\vert a_n \vert\,.
\end{equation}
The weight of the product of harmonic sums is equal to the sum of their weights.

For twist-two operators, there is an additional simplification, thanks to the so-called generalized Gribov-Lipatov reciprocity~\cite{Gribov:1972ri,Gribov:1972rt,Dokshitzer:2005bf,Dokshitzer:2006nm}, which reflects the symmetry of the underlying process under the crossing of channels. A consequence of this property is that the harmonic sums can enter the anomalous dimension only in the form of special combinations, which satisfy the property mentioned above. In practice, this allows one to restrict the choice of possible basis functions to the smaller number of so-called binomial harmonic sums, which are defined as~\cite{Vermaseren:1998uu}
\begin{equation}
{\mathbb S}_{a_1,\ldots,a_n}(M)=(-1)^M\sum_{j=1}^{M}(-1)^j\binom{M}{j}\binom{M+j}{j}S_{a_1,...,a_n}(j)\,.
\label{BinomialSums}
\end{equation}
They have only positive-integer indices, while their transcendentality is the same as for usual harmonic sums.

The second observation is that, in the expressions for the $j$-dependent anomalous dimensions already known, the coefficients in front of these sums are rather simple numbers, usually small integers.
In the general case, we thus obtain a system of Diophantine equations. If the number of equations is equal to the number of variables, then we can solve the system exactly, but, in this case, we need to know a lot of fixed values.
However, the system of Diophantine equations can be solved with help of a special method from number theory even if the number of equations is less than the number of variables. In this case, we use the Lenstra--Lenstra--Lovasz (LLL) algorithm~\cite{Lenstra82factoringpolynomials}, which allows one to reduce the matrix obtained from the system of Diophantine equations to a matrix the rows of which are the solution of the system with minimal Euclidean norm.

According to the maximal-transcendentality principle~\cite{Kotikov:2002ab}, anomalous dimensions of twist-two operators can only contain harmonic sums with the maximum weight allowed at the respective order of perturbation theory. The latter is equal to $2\ell-1$ at $\ell$-th order. In our case, of $\ell=4$, the basis for the four-loop universal anomalous dimension contains all possible binomial harmonic sums [see Eq.~\eqref{BinomialSums}] with weight $7$, and there are $2^{7-1}=64$ of them. This only applies to $\gamma_{{\mathrm{uni}},{\mathrm{np}},{\mathrm{rational}}}^{(3)}(\M)$ in Eq.~\eqref{HSZ5ResALL}. The weight of $\zeta_i$ is $i$, so that $\gamma_{{\mathrm{uni}},{\mathrm{np}},\zf}^{(3)}(\M)$ and $\gamma_{{\mathrm{uni}},{\mathrm{np}},\zt}^{(3)}(\M)$ in Eq.~\eqref{HSZ5ResALL} are of transcendentalities 4 and 2, respectively. The numbers of binomial sums in the respective bases are $8$ and $2$, respectively.

The general form of the $\zf$ part,
\begin{equation}\label{HSZ5ResS1}
\gamma_{{\mathrm{uni}},{\mathrm{np}},\zf}^{(3)}(\M) = -\,40\,{\mathbb S}_{1}^2(\M-2)\,,
\end{equation}
was determined a long time ago \cite{Velizhanin:2009gv} from the single input $\gamma_{\mathrm {uni,np}}^{(3)}(4)$ in Eq.~\eqref{guniM4}, which was sufficient to fix the two coefficients in the ansatz, and all the subsequent results for $\gamma_{\mathrm {uni,np}}^{(3)}(j)$, with $j=6,\ldots,18$, confirmed this assumption.
The general form of the $\zt$ part was obtained upon the derivation of $\gamma_{\mathrm {uni,np}}^{(3)}(12)$ in Eq.~\eqref{guniM12}, when five values were known, enough to fix the eight coefficients in the ansatz, and reads \cite{Kniehl:2020rip}
\begin{equation}\label{HSZ3ResS1}
\gamma_{{\mathrm{uni}},{\mathrm{np}},\zt}^{(3)}(\M) = 
8\, \Big(8\, {\HBS}_4-9\, {\HBS}_{1,3}-3\, {\HBS}_{2,2}-4\, {\HBS}_{3,1}+4\, {\HBS}_{1,1,2}+5\, {\HBS}_{1,2,1}-{\HBS}_{2,1,1}\Big),
\end{equation}
where $\BS_{\pmb{a}}=\BS_{\pmb{a}}(j-2)=\BS_{\pmb{a}}(M)$.
The results in Eqs.~\eqref{guniM14}--\eqref{guniM18} all satisfy Eq.~\eqref{HSZ3ResS1}.

To reconstruct the rational part $\gamma_{{\mathrm{uni}},{\mathrm{np}},{\mathrm{rational}}}^{(3)}(\M)$ in Eq.~\eqref{HSZ5ResALL}, we exploit three general properties which the non-planar anomalous dimension is supposed to satisfy. First of all, its large-$\M$ limit should not contain $\ln\M$ to a power higher than one~\cite{Korchemsky:1988si,Korchemsky:1992xv,Alday:2007mf} in the combinations of the basis functions. Specifically, this constraint demands the absence of the six terms $\ln^7\!\M$, $\z2\ln^5\!\M$, $\z3\ln^4\!\M$, $\z4\ln^3\!\M$, $\z2\z3\ln^2\!\M$, and $\z5\ln^2\!\M$.
Moreover, the result for the four-loop cusp anomalous dimension~\cite{Henn:2019swt,Huber:2019fxe,vonManteuffel:2020vjv} fixes the coefficient of the term proportional to $\ln\M$.
Specifically, this determines the two coefficients in the front of $\z3^2\ln\M$ and $\z6\ln\M$.
The second property comes from the Balitsky--Fadin--Kuraev--Lipatov (BFKL) equation~\cite{Lipatov:1976zz,Kuraev:1977fs,Balitsky:1978ic}, which is exactly known at the first two orders, in the leading-logarithmic approximation (LLA) and the next-to-leading-logarithmic approximation (NLLA)~\cite{Kotikov:2002ab,Fadin:1998py,Kotikov:2000pm}. This implies that, being analytically continued into $\M=1+\omega$ (or $M=-1+\omega$), the universal anomalous dimension of the twist-two operators in ${\mathcal N}=4$ SYM theory should not contain poles in $\omega$ less than $1/\omega^{\ell-1}$ at $\ell$ loops, which, in our case, leaves the poles $1/\omega^k$ with $k=3,\ldots,7$ and so impose five constraints.
The third constraint is due to the double-logarithmic equation~\cite{Kirschner:1982qf,Kirschner:1983di} and implies that, analytically continued to $\M=0+\omega$ (or $M=-2+\omega$), the universal anomalous dimension of twist-two operators in ${\mathcal N}=4$ SYM theory should not contain the highest pole in $\omega$, which excludes the pole $1/\omega^7$ in our case. For the analytic continuation of the harmonic sums, we rely on our database~\cite{Velizhanin:2020avm}, generated with the help of the
\texttt{FORM}~\cite{Vermaseren:2000nd} packages \texttt{harmpol}~\cite{Remiddi:1999ew} and \texttt{summer}~\cite{Vermaseren:1998uu}, and the collection \texttt{DATAMINE}~\cite{Blumlein:2009cf} of relations between alternating multiple zeta values. So, we have $8+(6+2)+5+1=22$ equations for $64$ variables.

To gain confidence in our reconstruction procedure, we apply it to recover the well-known result for the rational part of the planar four-loop anomalous dimension that respects reciprocity (RR) \cite{Kotikov:2007cy,Bajnok:2008qj}, which, in the basis of the binomial harmonic sums, reads
\begin{eqnarray}
\frac{\gamma_{\mathbb{RR},\mathrm {uni},\mathrm {pl},\mathrm {rat}}^{4-\mathrm{loop}}}{32}&=& 
-2 \BS_{1,2,4}
+2 \BS_{1,5,1}
+4 \BS_{2,2,3}
+\BS_{2,3,2}
-5 \BS_{2,4,1}
+2 \BS_{3,1,3}
-3 \BS_{3,2,2}
-\BS_{3,3,1}
\nonumber\\&&
{}+4 \BS_{4,1,2}
-4 \BS_{4,2,1}
+2 \BS_{5,1,1}
+4 \BS_{1,1,1,4}
-6 \BS_{1,1,2,3}
-2 \BS_{1,1,3,2}
+4 \BS_{1,1,4,1}
\nonumber\\&&
{}-2 \BS_{1,2,1,3}
+9 \BS_{1,2,2,2}
-9 \BS_{1,2,3,1}
+3 \BS_{1,3,1,2}
-5 \BS_{1,3,2,1}
+4 \BS_{1,4,1,1}
-2 \BS_{2,1,1,3}
\nonumber\\&&
{}+5 \BS_{2,1,2,2}
-5 \BS_{2,1,3,1}
+8 \BS_{2,2,2,1}
-6 \BS_{2,3,1,1}
+3 \BS_{3,1,1,2}
-\BS_{3,1,2,1}
-4 \BS_{3,2,1,1}
\nonumber\\&&
{}+2 \BS_{4,1,1,1}
-4 \BS_{1,1,1,2,2}
+6 \BS_{1,1,1,3,1}
+2 \BS_{1,1,2,1,2}
-8 \BS_{1,1,2,2,1}
+4 \BS_{1,1,3,1,1}
\nonumber\\&&
{}+2 \BS_{1,2,1,1,2}
-4 \BS_{1,2,2,1,1}
+2 \BS_{1,3,1,1,1}
+2 \BS_{2,1,1,1,2}
-2 \BS_{2,2,1,1,1}
\end{eqnarray}
In fact, we find that the 22 equations described above can correctly fix all the coefficients of the 64 binomial harmonic sums using the \texttt{fplll} lattice reduction library~\cite{fplll}, in which the LLL algorithm and other similar algorithms are implemented. It is, therefore, reasonable to expect that our reconstruction procedure will also work in the case at hand. We find
\begin{eqnarray}
\frac{\gamma_{{\mathrm{uni}},{\mathrm{np}},{\mathrm{rat}}}^{(3)}(\M)}{4}&=& 
2 \BS_{5,2}
-2 \BS_{4,3}
-4 \BS_{1,2,4}
-2 \BS_{1,3,3}
+2 \BS_{1,4,2}
+4 \BS_{1,5,1}
-6 \BS_{2,2,3}
+2 \BS_{2,3,2}
\nonumber\\&&
{}+4 \BS_{2,4,1}
+12 \BS_{3,1,3}
-6 \BS_{3,2,2}
-2 \BS_{3,3,1}
+2 \BS_{4,1,2}
+6 \BS_{4,2,1}
-12 \BS_{5,1,1}
\nonumber\\&&
{}+8 \BS_{1,1,1,4}
+4 \BS_{1,1,2,3}
-8 \BS_{1,1,3,2}
-4 \BS_{1,1,4,1}
-\BS_{1,2,1,3}
+3 \BS_{1,2,2,2}
\nonumber\\&&
{}+6 \BS_{1,2,3,1}
-14 \BS_{1,3,1,2}
-3 \BS_{1,3,2,1}
+9 \BS_{1,4,1,1}
-3 \BS_{2,1,1,3}
-\BS_{2,1,2,2}
\nonumber\\&&
{}+\BS_{2,2,2,1}
+3 \BS_{2,3,1,1}
-8 \BS_{3,1,1,2}
-8 \BS_{3,1,2,1}
+12 \BS_{3,2,1,1}
+4 \BS_{4,1,1,1}
\nonumber\\&&
{}+4 \BS_{1,1,1,2,2}
-12 \BS_{1,1,1,3,1}
+4 \BS_{1,1,2,1,2}
-4 \BS_{1,1,2,2,1}
+8 \BS_{1,1,3,1,1}
\nonumber\\&&
{}+8 \BS_{1,2,1,1,2}
+2 \BS_{1,2,1,2,1}
-14 \BS_{1,2,2,1,1}
+4 \BS_{1,3,1,1,1}
+4 \BS_{2,1,1,1,2}
\nonumber\\&&
{}+4 \BS_{2,1,1,2,1}
-4 \BS_{2,1,2,1,1}
-4 \BS_{2,2,1,1,1}
\,,\label{NPADFL}
\end{eqnarray}
where $\BS_{\pmb{a}}=\BS_{\pmb{a}}(j-2)=\BS_{\pmb{a}}(M)$.
Technical details of our evaluation may be found in the Appendix. Although we believe that our result in Eq.~\eqref{NPADFL} is correct, it needs to be checked, either directly by calculating the next moment, for $j=20$, or indirectly by considering some limiting cases. Below, we present our predictions for some limiting cases of $\gamma_{{\mathrm{uni}},{\mathrm{np}}}^{(3)}(\M)$, evaluated with Eqs.~\eqref{HSZ5ResS1}, \eqref{HSZ3ResS1}, and \eqref{NPADFL}.

First of all, we consider the large-$\M$ limit of $\gamma_{{\mathrm{uni}},{\mathrm{np}}}^{(3)}(\M)$, which yields
\begin{equation}
\gamma_{\mathrm{uni},\mathrm{np}}^{(3)\ }\overset{\M\to\infty}{=}
\left(-24 \z3^2-62 \z6\right)\ln \M +4 \z4 \z3-20 \z2\z5-175 {\z7}\,.\label{largejlimit}
\end{equation}
We observe that the $j$-independent term in Eq.~\eqref{largejlimit}, which takes the numerical value $-205.37$, nicely agrees with the value $-B_{\mathrm{np}}^{(4)}/48=(-207.0\pm3.0)$ we obtained in Ref.~\cite{Kniehl:2020rip}.

The analytic continuation of $\gamma_{{\mathrm{uni}},{\mathrm{np}}}^{(3)}(\M)$ to the BFKL value $M=-1+\omega$ yields\footnote{We have used $M$ instead of $j=M+2$ to comply with the standard convention in $\mathcal N=4$ SYM theory.}
\begin{equation}
\gamma_{\mathrm{uni},\mathrm{np}}^{(3)}\overset{M=-1+\omega}{=}
-\frac{192 }{\omega^2}{\z2} {\z3}
+\frac{4 }{\omega}
\left(
31 {\z6}
-2 {\z3}^2
\right)
+2 \left(
20 {\z2} {\z5}
+280 {\z3} {\z4}
+69 {\z7}
\right)+\mathcal{O}(\omega)\,.\qquad\label{BFKLlimit}
\end{equation}
This implies that the BFKL equation receives a non-planar contribution in the next-to-next-to-leading logarithmic approximation (NNLLA), which is the third order of perturbation theory.

In the double-logarithmic limit, $M=-2+\omega$, we have
\begin{eqnarray}
\gamma_{\mathrm{uni},\mathrm{np}}^{(3)}&\overset{M=-2+\omega}{=}&
\frac{192 {\z2}}{\omega^5}
-\frac{384 {\z2}}{\omega^4}
-\frac{48}{\omega^3}\left(4 {\z2}+15 {\z4}\right)
+\frac{288}{\omega^2}\left(4 {\z2} {\z3}+5 {\z4}\right)\nonumber\\&&
{}+\frac{4 }{\omega}
\left(
144 {\z2}
-1728 {\z2} {\z3}
-24 {\z3}^2
-540 {\z4}
+60 {\z5}
-1367 {\z6}
\right)\nonumber\\&&
{}+4 \left(
96 {\z2}
-288 {\z2} {\z3}
+722 {\z2} {\z5}
+36 {\z3}^2
-814 {\z3} {\z4}
+799 {\z6}
-56 {\z7}
\right)+\mathcal{O}(\omega)\,.
\nonumber\\
&&\label{DLlimit}
\end{eqnarray}
As in BFKL case, the double-logarithmic equation receives a non-planar contribution in the third order of perturbation theory.

Another limit of interest is $M=0+\omega$, which was taken for the planar case in Ref.~\cite{Basso:2011rs,Gromov:2014bva}.
In this limit, we have
\begin{eqnarray}
\gamma_{\mathrm{uni},\mathrm{np}}^{(3)}&\overset{M=0+\omega}{=}&
\omega\left(
-72\z2\z3^2
-\frac{1870}{3}\,\z8
\right)
+\omega^2\left(
165\z9
-320\z4\z5
+462\z3\z6
+490\z2\z7
\right)\nonumber\\&&
{}+\omega^3\left(
\frac{720706}{1583}\z3\z7
-122\z3^2\z4
-\frac{87788}{57}\z2\z3\z5
-\frac{86900}{1583}\z5^2
+\frac{13564298077}{10827720}\zz10\right.
\nonumber\\&&
{}+\left.\frac{25344}{1583}\hh73
-\frac{709632}{1583}\hh91
-\frac{75008}{513}\hh53\z2
+\frac{375040}{171}\hh71\z2
\right)
+\mathcal{O}(\omega^4)\,,\quad\label{M0expansion}
\end{eqnarray}
where $\mathrm{h}_{ab}$ are the special numbers related with the multiple zeta's (see Ref.~\cite{Blumlein:2009cf}).

Finally, also the value of $\gamma_{{\mathrm{uni}},{\mathrm{np}}}^{(3)}(\M)$ at $M=1$ is of special interest because it is related to the single-magnon operator at $\beta =1/2$ of the SU(2)$_\beta$ spin chain of marginally $\beta$-deformed $\mathcal {N}= 4$ SYM theory~\cite{Gunnesson:2009nn,Arutyunov:2010gu}. From Eq.~\eqref{NPADFL}, we obtain
\begin{equation}
\gamma_{\mathrm{uni},\mathrm{np}}^{(3)}(3)=-160\z5\,.\label{betadeformed}
\end{equation}
It is surprising that, in the non-planar case, only the most transcendental contribution survives, which can even be found by computing the non-planar Konishi contribution in Eq.~\eqref{guniM4} because, for the reconstruction of the general expression of the most transcendental part, only one fixed value is necessary to find the coefficient of $\BS_1^2$ in Eq.~\eqref{HSZ5ResS1}.

\section{Conclusion}

We constructed the general analytic expression for the non-planar contribution to the four-loop universal anomalous dimensions in $\mathcal{N}=4$ SYM theory, given in Eq.~\eqref{NPADFL}, using the first eight moments in Eqs.~\eqref{guniM4}--\eqref{guniM8} \cite{Velizhanin:2009gv,Velizhanin:2010ey,Velizhanin:2014zla}
and~\eqref{guniM10}--\eqref{guniM18} \cite{Kniehl:2020rip} together with information on the large-$j$~\cite{Huber:2019fxe,Henn:2019swt,vonManteuffel:2020vjv}, BFKL \cite{Kotikov:2002ab,Fadin:1998py,Kotikov:2000pm}, and double-logarithmic \cite{Kirschner:1982qf,Kirschner:1983di} limits, assuming that the ansatz should include only binomial harmonic sums [see Eq.~\eqref{BinomialSums}] in its coefficients, and solving the resulting system of Diophantine equations with the help of number theory. We also provided analytic results for various interesting limits of our general expression, including the
large-$j$ limit in Eq.~\eqref{largejlimit}, the BFKL limit in Eq.~\eqref{BFKLlimit}, the double-logarithmic limit in Eq.~\eqref{DLlimit}, the expansion about $M=0$ in Eq.~\eqref{M0expansion}, and the result for $M=1$ in Eq.~\eqref{betadeformed}. The latter is particularly interesting, as it represents a very simple expression for the non-planar contribution to the anomalous dimension of the single-magnon operator at $\beta=1/2$ of the SU(2)$_\beta$ spin chain of marginally $\beta$-deformed $\mathcal{N}=4$ SYM theory.
It will be tantalizing to test these conjectures in the future by independent calculations.

\subsection*{Acknowledgments}

We would like to thank V.~S.~Fadin for useful discussions.
Our computations were performed in part with resources provided by the PIK Data Centre in PNPI NRC ``Kurchatov Institute.''
The research of B.A.K. was supported in part by BMBF Grant No.\ 05H18GUCC1 and
DFG Grants No.\ KN~365/13-1 and No.\ KN~365/14-1.
The research of V.N.V. was supported in part by RFBR Grants No.\ 16-02-00943-a, No.\ 16-02-01143-a, and No.\ 19-02-00983-a and a Marie Curie International Incoming Fellowship within the Seventh European Community Framework Programme under Grant No.\ PIIF-GA-2012-331484.


\section{Appendix}

Here, we provide more details on the reconstruction procedure and the result, which we obtain with the help of number theory.
As discussed in Section~\ref{Reconstruction}, the ansatz for the general result consists of a linear combination of the binomial harmonic sums [see Eq.~\eqref{BinomialSums}] of weight 7, of which there are $2^{7-1}=64$ in total,
\begin{eqnarray}
\mathrm{Ansatz}_{\mathrm {Rat}}
&=&\Big\{\BS_{7}, \BS_{1, 6}, \BS_{2, 5}, \BS_{3, 4}, \BS_{4, 3}, 
 \BS_{5, 2}, \BS_{6, 1}, \BS_{1, 1, 5}, \BS_{1, 2, 4}, 
 \BS_{1, 3, 3}, \BS_{1, 4, 2}, \BS_{1, 5, 1}, \BS_{2, 1, 4}, \nonumber\\&&
 \BS_{2, 2, 3}, \BS_{2, 3, 2}, \BS_{2, 4, 1}, \BS_{3, 1, 3}, 
 \BS_{3, 2, 2}, \BS_{3, 3, 1}, \BS_{4, 1, 2}, \BS_{4, 2, 1}, 
 \BS_{5, 1, 1}, \BS_{1, 1, 1, 4}, \BS_{1, 1, 2, 3},  \nonumber\\&&
 \BS_{1, 1, 3, 2}, \BS_{1, 1, 4, 1}, \BS_{1, 2, 1, 3}, 
 \BS_{1, 2, 2, 2}, \BS_{1, 2, 3, 1}, \BS_{1, 3, 1, 2}, 
 \BS_{1, 3, 2, 1}, \BS_{1, 4, 1, 1}, \BS_{2, 1, 1, 3},  \nonumber\\&&
 \BS_{2, 1, 2, 2}, \BS_{2, 1, 3, 1}, \BS_{2, 2, 1, 2}, 
 \BS_{2, 2, 2, 1}, \BS_{2, 3, 1, 1}, \BS_{3, 1, 1, 2}, 
 \BS_{3, 1, 2, 1}, \BS_{3, 2, 1, 1}, \BS_{4, 1, 1, 1},  \nonumber\\&&
 \BS_{1, 1, 1, 2, 2}, \BS_{1, 1, 1, 3, 1}, \BS_{1, 1, 2, 1, 2}, 
 \BS_{1, 1, 2, 2, 1}, \BS_{1, 1, 3, 1, 1}, \BS_{1, 2, 1, 1, 2}, 
 \BS_{1, 2, 1, 2, 1}, \BS_{1, 2, 2, 1, 1},  \nonumber\\&&
 \BS_{1, 3, 1, 1, 1}, 
 \BS_{2, 1, 1, 1, 2}, \BS_{2, 1, 1, 2, 1}, \BS_{2, 1, 2, 1, 1}, 
 \BS_{2, 2, 1, 1, 1}, \BS_{3, 1, 1, 1, 1}, \BS_{1, 1, 1, 2, 1, 1}, 
 \BS_{1, 1, 2, 1, 1, 1},  \nonumber\\&&
 \BS_{1, 2, 1, 1, 1, 1}, 
 \BS_{2, 1, 1, 1, 1, 1}, \BS_{1, 1, 1, 1, 3}, 
 \BS_{1, 1, 1, 1, 2, 1}, 
 \BS_{1, 1, 1, 1, 1, 2}, 
 \BS_{1, 1, 1, 1, 1, 1, 1}\Big\}\,.\label{Ansatz}
\end{eqnarray}
We have $22$ constraints, which are described in Section~\ref{Reconstruction}, and construct from these equations a matrix of dimension $22\times(64+1)$, where the last column is multiplied by the factor $2^{-3}$, which we expect to be a common factor of the desired expression. Then we append the transpose of this matrix, multiplied by any large number, $8^{64}$ in our case, to the right of the unit matrix of dimension $65\times65$. The resulting matrix, which is of dimension $65\times(65+22)$, is then translated into a form which can be read by the \texttt{fplll} library~\cite{fplll}, in which the LLL algorithm and further modifications of it are implemented. We use the so-called Block-- Korkine--Zolotarev (BKZ) method,\footnote{See Refs.~\cite{Korkine_A_Zolotareff_G,SCHNORR1987201,Hanrot:2008lll} for more details.}
with block size set to $25$ using option \texttt{-b} and precision set to 555 using options \texttt{-mpfr -p}. The library \texttt{fplll} produces a new matrix in which the lines are sorted according to their Euclidean norm. Upon the execution of \texttt{fplll}, we have a new matrix, which contains the solutions of the original non-uniform Diophantine system of equations and also the solutions of the corresponding uniform system of equations---the number of equations is less than the number of variables, and there are combinations of variables which satisfy the system of uniform equations. The first line of this matrix is the solution of the uniform equation, with 0 in position 65,\footnote{We suppress
$22$ zeros at the end of the line, as this is a general property of the solutions.}
\begin{eqnarray}
\mathrm{fplll}_1&=&\big\{0, 0, -3, 1, -6, -3, 3, -9, -3, 1, -2, 2, 2, 8, 1, -3, 1, -5, -7, 9, 
9, 0, 10, 9, 3, 5, 0, \nonumber\\&&
-3, 3, -1, 3, 0, 1, 7, -1, -12, 0, 5, -1, 1, 
11, -12, -2, -10, -7, -7, 1, -6, 1, 3, \nonumber\\&&
-3, 10, -11, -8, 4, -1, 2, 5, 
5, 0, 0, 0, 0, 0, 0\big\}\,,\label{line1}
\end{eqnarray}  
while the second line gives the desired solution of the non-uniform equation, with 2 in position 65,
\begin{eqnarray}
\mathrm{fplll}_2&=&\big\{0, 0, 0, 0, 2, -2, 0, 0, 4, 2, -2, -4, 0, 6, -2, -4, -12, 6, 2, -2, 
-6, 12, -8, -4, 8, \nonumber\\&&
4, 1, -3, -6, 14, 3, -9, 3, 1, 0, 0, -1, -3, 8, 8, 
-12, -4, -4, 12, -4, 4, -8, -8, \nonumber\\&&
-2, 14, -4, -4, -4, 4, 4, 0, 0, 0, 0, 
0, 0, 0, 0, 0, 2
\big\}\,.\label{line2}
\end{eqnarray}  
The next possible solution of the non-uniform system is on line 20. So, there is reason to assume that the solution in Eq.~\eqref{line2} is correct because
\begin{itemize}
\item it is isolated from other solutions;
\item it contains many zeros;
\item it contains the same coefficients, up to a sign, for the most complicated binomial harmonic sums in ansatz~\eqref{Ansatz}, i.e.\ those at the end of the ansatz;
\item it exhibits a rather uniform structure.
\end{itemize}
In principle, any line of the obtained matrix can be added to line 2 accommodating our conjectured solution.
However, as can be seen by adding Eqs.~\eqref{line1} and \eqref{line2}, we then loose the above-mentioned properties of the desired result, which we expect from our observations made when computing anomalous dimensions.  

\bibliographystyle{plain}


\end{document}